\documentclass[aps,pra,reprint,groupedaddress]{revtex4-1}
\usepackage{xcolor}
\usepackage{hyperref}
\hypersetup{colorlinks=false,linkbordercolor=red,linkcolor=green,pdfborderstyle={/S/U/W 1}}
\usepackage[UKenglish]{babel}
\usepackage[]{graphicx, amsfonts, amsmath, amssymb, amstext, latexsym, float, color, hyperref, mathtools}

\newcommand{\ket}[1]{|#1\rangle}	
\newcommand{\bra}[1]{\langle#1|}

\newcommand{\sinc}{\text{sinc}}
\newcommand{\sinctmu}{\text{sinc}~2\mu}
\newcommand{\sincmu}{\text{sinc}~\mu}
\newcommand{\be}{\begin{equation}}
\newcommand{\ee}{\end{equation}}

\newcommand{\mx}[1]{\begin{pmatrix}#1\end{pmatrix}}

\begin{document}

\title{Attaining the quantum limit of passive imaging}

\author{Hari Krovi}
\thanks{Email of corresponding author: hkrovi@bbn.com. This work was partially presented by Saikat Guha as an invited talk at the Single Photon Workshop (SPW) 2013, held at ORNL.}
\affiliation{Quantum Information Processing group, Raytheon BBN Technologies, Cambridge, MA 02138, USA}
\author{Saikat Guha}
\affiliation{Quantum Information Processing group, Raytheon BBN Technologies, Cambridge, MA 02138, USA}
\author{Jeffrey H. Shapiro}
\affiliation{Research Laboratory of Electronics, Department of Electrical Engineering and Computer Science, Massachusetts Institute of Technology, Cambridge, MA 02139, USA}

\begin{abstract}
We consider the problem, where a camera is tasked with determining one of two hypotheses: first with an incoherently-radiating quasi-monochromatic point source and the second with two identical closely spaced point sources. We are given that the total number of photons collected over an integration time is assumed to be the same under either hypothesis. For the one-source hypothesis, the source is taken to be on-axis along the line of sight and for the two-source hypothesis, we give ourselves the prior knowledge of the angular separation of the sources, and they are assumed to be identical and located symmetrically off-axis. This problem was studied by Helstrom in 1973, who evaluated the probability of error achievable using a sub-optimal optical measurement, with an unspecified structured realization. In this paper, we evaluate the quantum Chernoff bound, a lower bound on the minimum probability of error achievable by any physically-realizable receiver, which is exponentially tight in the regime that the integration time is high. We give an explicit structured receiver that separates three orthogonal spatial modes of the aperture field followed by quantum-noise-limited time-resolved photon measurement and show that this achieves the quantum Chernoff bound. In other words, the classical Chernoff bound of our mode-resolved detector exactly matches the quantum Chernoff bound for this problem. Finally, we evaluate the classical Chernoff bound on the error probability achievable using an ideal focal plane array---a signal shot-noise limited continuum photon-detection receiver with infinitely many infinitesimally-tiny pixels---and quantify its performance gap with the quantum limit. 
\end{abstract}
\maketitle

Numerous example problems in optical communications and sensing are known where the fundamental performance limits predicted by quantum mechanics are superior to the ultimate performance limits associated with conventional optical receivers, even when these receivers are assumed to be operating at their respective quantum-noise-limited performance. Choosing a specific optical receiver design to detect the received optical field---{\em detection} referring to the act of converting the (quantum) optical field to a noisy (classical) electrical signal---results in a probabilistic description of the problem. This probabilistic noisy-channel description takes the form of an input-output transition probability matrix $p_{Y|X}(y|x)$ for a communication problem where the receiver's task is to decode messages encoded in a long sequence of the input variable $X$ from a corresponding sequence of the channel's output $Y$, at the minimum probability of error. In an imaging problem, the probabilistic description is the channel output $p_{Y|{\Theta}}(y|\theta)$ where the receiver's task is to estimate a scene parameter $\theta$ with minimum error (e.g., mean-squared error) from a sequence of the channel output $Y$. Once a specific optical receiver is chosen, one can calculate the associated probabilistic channel description as above, after which (classical) information and estimation theoretic tools can be used to find the optimal performance achievable for the task at hand, e.g., the maximum reliable communication rate or the minimum mean-squared error in estimation. Quantum information and detection theory gives us mathematical tools to directly compute the optimal achievable performance optimized over absolutely any possible receiver design. By definition, such a performance limit must exceed the performance limit associated with a specific receiver design. Unfortunately however, even through quantum tools give us the fundamental performance limits, they usually do not translate readily into prescriptions of structured receiver designs that can actually attain the quantum-limited performance. Finding the structured design of a quantum-optimal receiver often requires creative bottom-up design. Most such non-standard optical receivers that achieve (or approach) quantum-limited performance of an imaging or communication task, involve some form of an all-optical transformation of the received optical field prior to detection, and may even involve incremental-detection-induced electro-optic feedback that modulates the pre-detection all-optical transformation of the received field while it gets detected. One classic example of that is Dolinar's receiver design to discriminate between two apriori-known laser-light waveforms, which exactly attains the minimum error probability permissible by quantum mechanics, and thereby outperforms both shot-noise-limited homodyne detection and direct detection receivers~\cite{Dol73}. The Dolinar receiver is one of many examples where even though the light involved is {\em classical} (one with a proper $P$-function description, i.e., can be expressed as a statistical mixture of coherent states of the radiation field), the optimal performance is {\em not} achieved by one of the conventional ways of detecting light, viz., homodyne, heterodyne, or direct detection.

\begin{figure}
\centering
\includegraphics[width=\columnwidth]{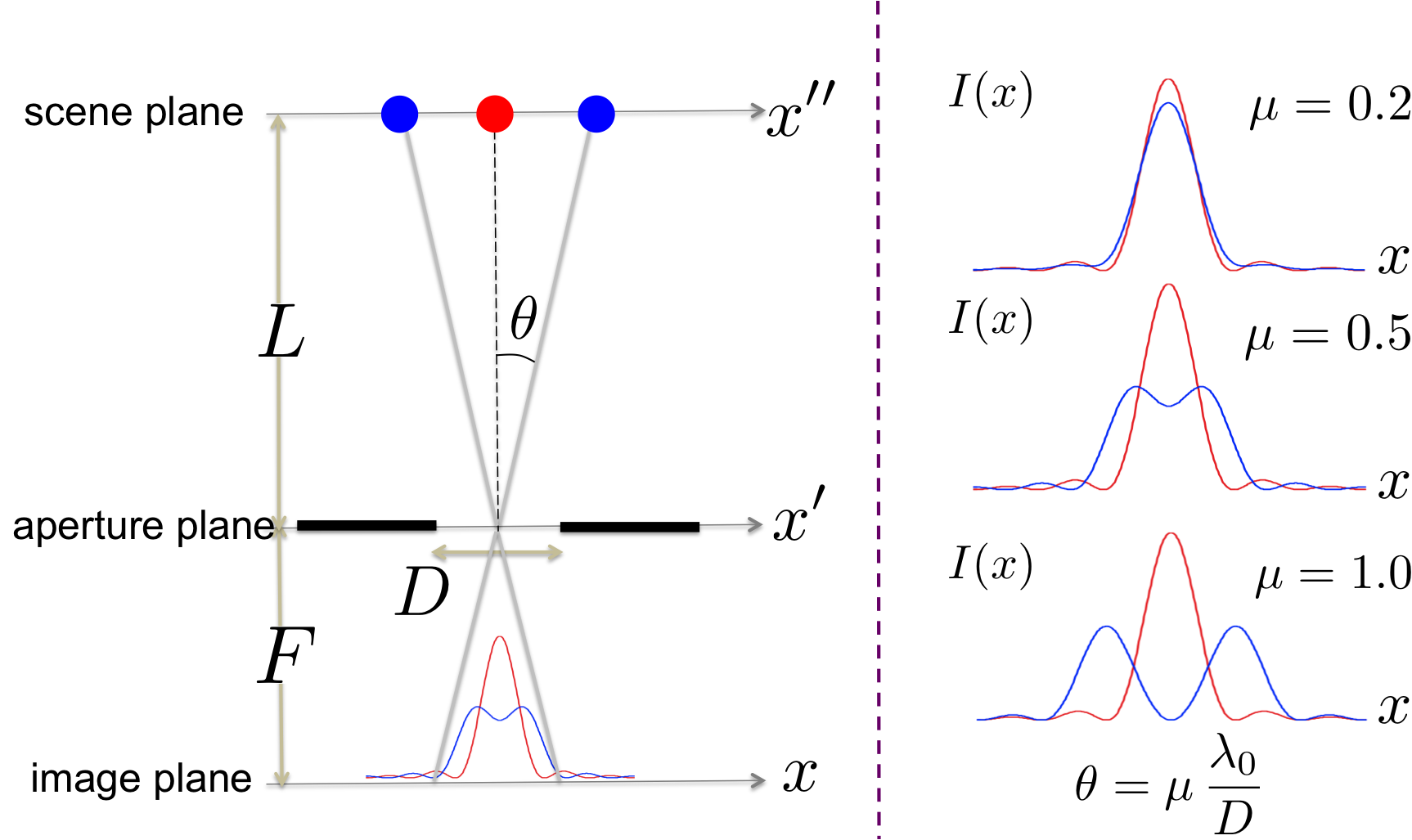}
\caption{(a) Discriminating one from two point sources incoherently radiating at center-wavelength $\lambda_0$. (b) The photon-unit image-plane intensity $I(x)$ shown for one source (red) and two sources (blue) for different values of angular separation of the two sources. The total photon number $\int_{-\infty}^{\infty} I(x) dx \equiv N$ is taken to be the same under both hypotheses.}
\label{fig:setup}
\end{figure}
In this paper, we evaluate the fundamental performance limit of an optical receiver's ability to discriminate one from two identical closely-spaced incoherently-radiating point sources in the sense of minimizing the probability of making an error, we quantify the performance gap to that achieved by a conventional image-plane focal plane array operating at its quantum-noise limit, and design a structured receiver that can attain the quantum-limited performance. 

A schematic of the problem setup is shown in Fig.~\ref{fig:setup}(a). The point sources radiate quasi-monochromatic incoherent light with a center frequency $\omega_0$, center wavelength $\lambda_0 \equiv \frac{2 \pi c}{\omega_0}$ ($c$ is the speed of light), and a spectral density of width $W$ (in Hz) that is much less than $\omega_0$. For the one-source hypothesis ($H_0$), the point source is assumed to be on light of sight (LoS), whereas for the two-source hypothesis ($H_1$), we assume the half-angular separation $\theta$ to be known a priori and that the sources are symmetrically disposed about the LoS and are equally radiating. We set up the problem in one spatial dimension, with $x \in (-\infty, \infty)$ denoting the spatial coordinate in the image plane (the focal plane of the conventional camera), $x^\prime \in (-\infty, \infty)$ denoting the spatial coordinate in the aperture plane, and $x^{\prime\prime} \in (-\infty, \infty)$ denoting the coordinate in the plane of the scene. We assume a $1$-D hard-aperture pupil of diameter (length) $D$, whose aperture function is given by $A(x^\prime) = {\rm rect}(x^\prime/D)$. Assuming Fraunhofer propagation over an $L$-m vacuum path from the scene to the receiver aperture's entrance pupil (which is accurate if the largest spatial dimension of interest in the scene plane $\ll \sqrt{\lambda_0L}$), and a lens at the entrance pupil perfectly focusing the image in an image plane located $F$-m behind the aperture (see Fig.~\ref{fig:setup}(a)), the image plane photons/m-unit intensity is given by:
\be
I(x) = \left\{
\begin{array}{ll}
{l N} |s(x)|^2 \equiv I_0(x), & H_0 \\
\frac{lN}{2} \left[|s(x - \theta F)|^2 + |s(x + \theta F)|^2\right] \equiv I_1(x), & H_1
\end{array}
\right. \nonumber
\ee
where $l = \lambda_0 F/D$ and $s(x) = \int_{-\infty}^\infty A(\lambda_0 F f)e^{j2 \pi fx}df$ is the point spread function (PSF). The total mean photon number collected over the receiver's integration time $T$, $\int_{-\infty}^{\infty} I(x) dx \equiv N$ is taken to be the same under both hypotheses. In Fig.~\ref{fig:setup}(b), we plot the image-plane intensity $I(x)$ for the one target hypothesis $H_0$ and for the two target hypothesis $H_1$, for $\mu \equiv \theta D/\lambda_0 = 0.2, 0.5$ and $1.0$. The latter is an incoherent sum of two symmetrically-shifted photon-unit squared-magnitude PSFs. 

This problem was first considered by Helstrom in 1973~\cite{Hel73}, who argued---via decomposing the aperture-collected field into mutually orthogonal spatial and temporal modes---that when $WT \gg 1$, the collected field spans roughly $M \approx WT$ statistically independent temporal modes, and that only the following three linearly-independent (but mutually non-orthogonal) spatial modes of the aperture field,
\begin{eqnarray}
\xi_1(x^\prime)&=&\left({1}/{\sqrt{D}}\right)\text{rect}\left({x^\prime}/{D}\right), \label{eq:aperturemode1}\\
\xi_2(x^\prime)&=&\left({1}/{\sqrt{D}}\right)e^{jk_0\theta x}\,\text{rect}\left({x^\prime}/{D}\right), \,{\text {and}} \label{eq:aperturemode2}\\ 
\xi_3(x^\prime)&=&\left({1}/{\sqrt{D}}\right)e^{-jk_0\theta x}\,\text{rect}\left({x^\prime}/{D}\right), \label{eq:aperturemode3}
\end{eqnarray}
with $\int_{-\infty}^\infty |\xi_i(x^\prime)|^2 dx^\prime = 1$ for $i = 1, 2, 3$, and $k_0 = 2\pi/\lambda_0$ contain all information relevant to the hypothesis-test problem. Helstrom expressed the quantum description of the $3M$ spatio-temporal modes collected over the integration time $T$ as an $M$-fold tensor product of a $3$-mode classically-correlated zero-mean Gaussian state, each of mean photon number $N/M \equiv N_0 \ll 1$, and wrote the expression for the minimum error probability achievable for choosing between the two hypotheses assuming they occur with equal prior probability. He then evaluated the error probability attainable by a suboptimal measurement which detects the total photon number (over all $M$ temporal modes) of one spatial mode in ${\cal S} \equiv {\rm span}\left\{\xi_1(x^\prime), \xi_2(x^\prime), \xi_3(x^\prime)\right\}$, followed by a quantum-limited minimum-error-probability measurement to discriminate the quantum states of the remaining two modes in the orthogonal complement of the aforesaid span~\cite{Hel73}.

In this paper, we first evaluate, in Section~\ref{sec:DD}, the probability of error attained by an ideal (shot-noise-limited) continuum focal plane array in the image plane with a unity-fill-factor array of infinitely many infinitesimally tiny pixels. In Section~\ref{sec:quantum}, we evaluate the fundamental limit to the minimum error probability using tools from quantum detection theory. In Section~\ref{sec:structured}, we propose and analyze a receiver that separates three mutually-orthogonal modes in $\cal S$, detects them using three shot-noise-limited temporally-mode-resolved single-photon-sensitive detectors, and makes the final hypothesis test using the detection outcomes over all $M$ temporal modes (i.e., over the entire integration time $T$) of the three detectors. We show that this mode-resolved-detection receiver exactly achieves the quantum limited performance evaluated in Section~\ref{sec:quantum}. In Section~\ref{sec:general}, we conjecture that the quantum-optimal performance of a mode-resolved detection strategy is generally true for all incoherent optical imaging problems where the scene parameter(s) of interest are encoded in a classically-correlated thermal phase-insensitive multimode optical field, and show evidence in its favor. Recent work by Mankei Tsang and collaborators have lent significant evidence in favor of this conjecture~\cite{Tsa09,Tsa15,Tsa15b,Nai16,Tsa16_general}. In particular, Tsang {\em et al.}'s work has established---for the problem of estimating the angular separation between two point sources---that Rayleigh's criterion for imaging resolution is an artifact of the conventional philosophy of focusing the image on a focal plane array and measurement of the field's intensity profile, and that if one allows for a pre-detection mode sorting, the imager's performance (in estimating the angular separation between two point sources) is only a function of the total collected photon energy during the integration time, and is entirely independent of the actual angular separation, no matter how small it is~\cite{Tsa15b}. The fact that pre-detection optical-domain pre-processing can enhance an imager's performance has been long known and explored in the computational imaging community, and is usually termed {\em super-resolution}. We conclude the paper in Section~\ref{sec:conclusion} with discussion of some of the aforesaid related work, and future directions---in particular our thoughts on proving the general conjecture about the optimality of mode-resolved photo detection for imaging with incoherent light.

\section{Image plane direct detection}\label{sec:DD}

The PSF of the hard aperture $A(x^\prime)$ is readily calculated as, $s(x) = (D/F\lambda_0)\,{\rm sinc}\left(xD/F\lambda_0\right)$, with ${\rm sinc}(x) \equiv \frac{{\rm sin}(\pi x)}{\pi x}$. The image-plane intensity functions under the two hypotheses can be re-expressed as $I_i(y) = Np_i(y)$, $i=0, 1$, using a scaled coordinate $y = xD/\lambda_0F$, where $p_0(y)$ and $p_1(y)$ given by:
\begin{eqnarray}
p_0(y) &=& {\rm sinc}^2(y), {\text{and}} \label{eq:dist1}\\
p_1(y) &=& \frac12\left[{\rm sinc}^2(y-\mu) + {\rm sinc}^2(y+\mu)\right], 
\label{eq:dist2}
\end{eqnarray}
for $y \in {\mathbb R}$, with $\int_{-\infty}^{\infty}p_i(y) dy = 1$ for $i = 0, 1$. We will first consider a quantum-noise-limited continuum detector in the image plane, which mimics an idealized scenario of an array of infinitely-many infinitesimally-small shot-noise limited unity-quantum-efficiency detector pixels with unity fill factor. We will thereafter consider the effect of finite-width pixels.

\subsection{Continuum detector}\label{sec:DDcontinuum}
The optimal direct detection of the image-plane field is achieved via a hypothetical continuum detector, which generates a Poisson point process with rate function either $I_0(y)$ or $I_1(y)$ depending upon which of the two hypothesis is true~\footnote{The reason the direct-detection statistics is Poisson is that the total photon number in $M \gg 1$ statistically-independent modes each in a thermal state with mean photon number $N_0 \ll 1$ is close to Poisson with mean $N = MN_0$. The mean photon number per mode $N_0 \ll 1$ is true at optical frequencies.}. The minimum probability of error $P_{\rm DD}$, attained by a maximum-likelihood estimate on the detector output (which is a sequence of spatial positions in the image plane where photon ``clicks" are detected) is upper bounded by the Chernoff bound (see Appendix~\ref{app:CB}):
\begin{equation}
P_{\rm DD} \le \frac{1}{2}e^{-M \xi_{\rm DD}},
\label{eq:DD_CCB}
\end{equation}
where $M \approx WT$ is the total number of temporal modes detected over the integration time $T$, and the Chernoff exponent $\xi_{\rm DD}$ is given by \cite{Snyder}:
\be
\xi_{\rm DD} = N_0\, C_\mu,
\ee
where $C_\mu = \max\limits_{0 \le s \le 1} C_\mu(s)$ is the Chernoff exponent, with
\be
C_\mu(s)=\int_{-\infty}^\infty \left[sp_0(y) + (1-s)p_1(y) - p_0(y)^{s}p_1(y)^{1-s}\right] dy,
\ee
where $N_0 = \frac{N}{M}$ is the mean photon number in each temporal mode, which would be typically $\ll 1$ at optical frequencies. The upper bound in Eq.~\eqref{eq:DD_CCB} is asymptotically tight for $M \to \infty$, and hence is a good (yet conservative) estimate of the detector's actual performance. Note that the exponent in Eq.~\eqref{eq:DD_CCB}, $M \xi_{\rm DD} = MN_0C = NC_\mu$, where $N_0$ is a constant dependent upon the radiance of the source and the wavelength, whereas the number of temporal modes $M$ (hence $N$, the total collected photon number) is proportional to the integration time $T$. The normalized Chernoff exponent $C_\mu$ is a function only of $\mu$ (given the shape of the aperture's PSF), and characterizes how well the two hypotheses can be discriminated.

\begin{figure}
\centering
\includegraphics[width=0.8\columnwidth]{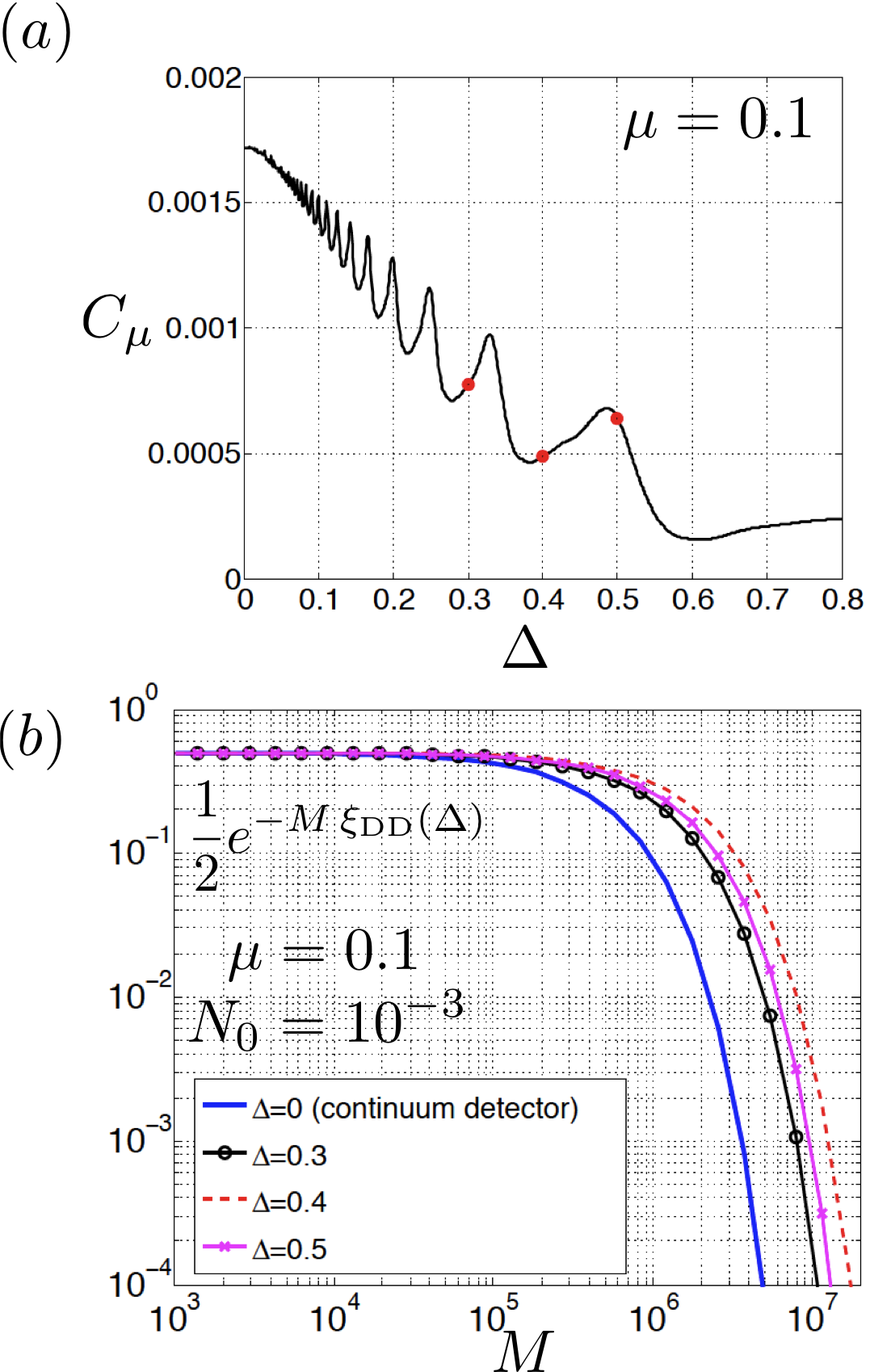}
\caption{(a) The plot of the Chernoff exponent $C_\mu(\Delta)$ of the pixelated FPA as a function of the pixel width $\Delta$, for normalized half-angular source separation $\mu = 0.1$; (b) The plot of the Chernoff (upper) bound on the probability of error achieved by a shot-noise-limited FPA as a function of $M \approx WT$, the total number of temporal modes collected over the integration time $T$, for $\mu=0.1$ and mean photon number per (temporal) mode $N_0 = 10^{-3}$. The figure shows the Chernoff bound plots for the continuum detector ($\Delta = 0$), and for pixelated detectors with $\Delta = 0.3, 0.4$ and $0.5$. The relative ordering of the plots are consistent with the values of $C_\mu(\Delta)$.}
\label{fig:FPA_pixelated_Pe}
\end{figure}

\subsection{Pixelated focal plane array}\label{sec:DDpixelated}

Let us now consider a pixelated focal plane array (FPA) with uniform pixel width $\Delta > 0$ with a pixel centered at $x=0$. Let us take the FPA to be of unity fill-factor, unity quantum efficiency, and shot-noise limited. The error probability achieved using this array $P_{\rm DD}(\Delta)$ is upper bounded by the Chernoff bound given by,
\begin{equation}
P_{\rm DD}(\Delta) \le \frac{1}{2}e^{-M \xi_{\rm DD}(\Delta)},
\label{eq:DD_CCB_FPA}
\end{equation}
where $\xi_{\rm DD}(\Delta) = N_0\, C_\mu(\Delta)$, and the normalized Chernoff exponent $C_\mu(\Delta) = \max\limits_{0 \le s \le 1} C_\mu(\Delta, s)$, with
\be
C_\mu(\Delta, s)=\sum_{n = -\infty}^\infty  (sq_0[n] + (1-s)q_1[n] - q_0[n]^s q_1[n]^{1-s}),
\ee
where $q_k[n] = \int_{(n-\frac12)\Delta}^{(n+\frac12)\Delta}p_k(y)dy$, for $k=0, 1$, $n \in {\mathbb Z}$, with $q_k[n]$ being the rate function of a space-discretized Poisson point process in the $n^{\rm th}$ detector pixel, which happens to add up to $1$, i.e., $\sum_{n=-\infty}^{\infty}q_k[n] = 1$, $k=0, 1$ (see Appendix~\ref{app:CB} for proof). The Chernoff exponent $C_\mu(\Delta)$ converges to $C_\mu$ as  $\Delta \to 0$. As expected, it (roughly) decreases as $\Delta$ (the pixelation) increases. However, as Fig.~\ref{fig:FPA_pixelated_Pe} shows, this degradation in performance due to increasing pixelation has an oscillatory nature, which is due to the oscillations in the sinc-function PSF of the hard aperture pupil.

\section{Quantum limit}\label{sec:quantum}

We will now evaluate the fundamental limit to the minimum probability of error attainable for this problem with no constraints whatsoever on how the aperture field is detected, as long as the detection method is allowed by physics. The quantum Chernoff bound (QCB) is a mathematical tool that lets us calculate an asymptotically-tight upper bound on this minimum error probability without making any reference to the actual optimal detection method. However, in order to invoke the QCB, we first need to express the (classical) aperture-collected field in terms of its quantum state representation. For this reason, the development in this Section will assume that the reader has some familiarity with quantum optics and detection theory. We refer the reader to~\cite{Hel76} for a comprehensive text on quantum estimation and detection theory with applications to quantum optics, and to~\cite{2008PirandolaLloyd} for a review of calculating the QCB for Gaussian state discrimination.

As stated in the introduction, the span $\cal S$ of the three spatial modes in Eqs.~\eqref{eq:aperturemode1},~\eqref{eq:aperturemode2} and~\eqref{eq:aperturemode3} contain all the relevant information in this imaging problem. Note that the phase sensitive moments are all zero. The optimal receiver may extract these three modes (over all $M$ statistically-independent temporal modes) before proceeding. One can calculate the inner products of the aforesaid functions as
\[
(\xi_1,\xi_2)=(\xi_1,\xi_3)=\sinc(\mu),\quad (\xi_2,\xi_3)=\sinc(2\mu)\,,
\]
with respect to the usual ${L}^2$-norm inner product, $(f,g)=\int_{-\infty}^\infty f(x^\prime) g^*(x^\prime) dx^\prime$. We use Gram-Schmidt orthonormalization to write down the following orthogonal modes that form a complete orthonormal basis for $\cal S$:
\begin{eqnarray}
\phi_1(x^\prime)&=&\xi_1(x^\prime),\\
\phi_2(x^\prime)&=&\frac{(\xi_2(x^\prime)-\xi_3(x^\prime))}{\sqrt{2(1-\sinc(2\mu))}}, {\text{and}}\\
\phi_3(x^\prime)&=&\frac{(-2\sinc(\mu)\xi_1(x^\prime)+\xi_2(x^\prime)+\xi_3(x^\prime))}{\sqrt{2(1+\sinc(2\mu)-2\sinc^2(\mu))}}.
\end{eqnarray}
The receiver's task therefore is to minimize the probability of error in discriminating $\rho_0^{\otimes M}$ and $\rho_1^{\otimes M}$, where $\rho_k$ is the joint quantum state of the three orthogonal spatial modes $(\phi_1, \phi_2, \phi_3)$. This minimum probability of error is upper bounded by the quantum Chernoff bound as:
\be
P_{\rm min} \le \frac{1}{2}e^{-M\xi_{Q}},
\label{eq:QCB}
\ee
with $\xi_Q = -\log Q_\mu$ is the quantum Chernoff exponent, with $Q_\mu = \min\limits_{0 \le s \le 1} Q_\mu(s)$, where
\be
Q_\mu(s) = {\rm Tr}\left(\rho_0^s\rho_1^{1-s}\right).
\ee
The quantum states $\rho_0$ and $\rho_1$ are zero-mean jointly-Gaussian states~\cite{Hel73}, and hence they are completely described by their second-order field moments, or equivalently by ${\boldsymbol V}^{(0)}$ and ${\boldsymbol V}^{(1)}$, the $6$-by-$6$ Wigner-distribution covariance matrices under the two hypotheses. 

Using the mode overlaps, we calculate the second-order field moments hypothesis $H_0$ as:
\be
\langle a_{\phi_1}^\dag a_{\phi_1}\rangle = N_0 \nonumber,
\ee
and,
\begin{eqnarray}
\langle a_{\phi_2}^\dag a_{\phi_2}\rangle &=& \langle a_{\phi_3}^\dag a_{\phi_3}\rangle \nonumber \\
&=& \langle a_{\phi_1}^\dag a_{\phi_2}\rangle \nonumber \\
&=& \langle a_{\phi_1}^\dag a_{\phi_3}\rangle \nonumber \\
&=& \langle a_{\phi_2}^\dag a_{\phi_3}\rangle = 0. \nonumber
\end{eqnarray}
All the phase-sensitive self and cross correlations (e.g., $\langle a_{\phi_2} a_{\phi_3}\rangle$, $\langle a_{\phi_1} a_{\phi_1}\rangle$) are zero. The $i,j$ entry of the first $3$-by-$3$ block of the covariance matrix ${\boldsymbol V}^{(0)}$ is given by $V^{(0)}_1(i,j)=\langle a_1^{\phi_i} a_1^{\phi_j}\rangle$, where $a_1^{\phi_i}=(1/2)(a_{\phi_i} + a_{\phi_i}^\dag)$. The off-diagonal $3$-by-$3$ blocks of the covariance matrix are given by $V^{(0)}_{12}(i,j)=(1/2)\langle a_1^{\phi_i} a_2^{\phi_j}+a_2^{\phi_i} a_1^{\phi_j}\rangle$ (similarly $V^{(0)}_{21}(i,j)$), where $a_2^{\phi_i}=(1/2i)(a_\phi - a_\phi^\dag)$ and $V^{(0)}_2(i,j)=\langle a_2^{\phi_i} a_2^{\phi_j}\rangle$. For hypothesis $H_0$, it can be seen that $V^{(0)}_1=V^{(0)}_2$ and $V^{(0)}_{12}=V^{(0)}_{21}=0$. Therefore, it suffices to write the first block $V^{(0)}_1$. It is simple to translate the field correlations to quadrature correlations, e.g.,
\[
\langle a_1^{\phi_1} a_1^{\phi_1}\rangle = \big{\langle} \left(\frac{a_{\phi_1}^\dag + a_{\phi_1}}{2}\right)^2\rangle = \frac{2N_0+1}{4}\,.
\]
Filling out the remainder of the entries, we get:
\be
V^{(0)}_1=\frac{1}{4}\mx{2N_0+1 & 0 & 0\\0 & 1 & 0\\0 & 0 & 1}\,.
\ee

Now for hypothesis $H_1$, all the phase-sensitive correlations are zero, and we have the following values for the various phase-insensitive second-order field moments:
\begin{align}
\langle a_{\phi_1}^\dag a_{\phi_1}\rangle =& 2N_0\sinc^2(\mu), \nonumber \\
\langle a_{\phi_2}^\dag a_{\phi_2}\rangle =&N_0(1-\sinc(2\mu)),\nonumber \\
\langle a_{\phi_3}^\dag a_{\phi_3}\rangle= &N_0(1+\sinc(2\mu)-2\sinc^2(\mu)), \nonumber \\
\langle a_{\phi_1}^\dag a_{\phi_2}\rangle =&\langle a_{\phi_2}^\dag a_{\phi_3}\rangle = 0, {\text{and}} \nonumber \\
\langle a_{\phi_1}^\dag a_{\phi_3}\rangle =& N_0\sinc(\mu)\sqrt{2(1+\sinc(2\mu)-2\sinc^2(\mu))} \nonumber \,.
\end{align}
It is a simple matter to translate the above field moments to quadrature moments, and obtain the following for the first $3$-by-$3$ block of ${\boldsymbol V}^{(1)}$:
\[
V^{(1)}_1=\frac{1}{4}\mx{N_0A^2+1 & 0 & N_0AB\\0 & 2N_0(1-\sinc(2\mu))+1 & 0\\N_0AB & 0 &N_0B^2+1}\,,
\]
where we have two $\mu$-dependent constants $A=2\sinc(\mu)$, and $B=\sqrt{2(1+\sinc(2\mu)-2\sinc^2(\mu))}$. We again have $V^{(1)}_2=V^{(1)}_1$, and $V^{(1)}_{12}=V^{(1)}_{21}=0$.

\begin{figure}
\centering
\includegraphics[width=\columnwidth]{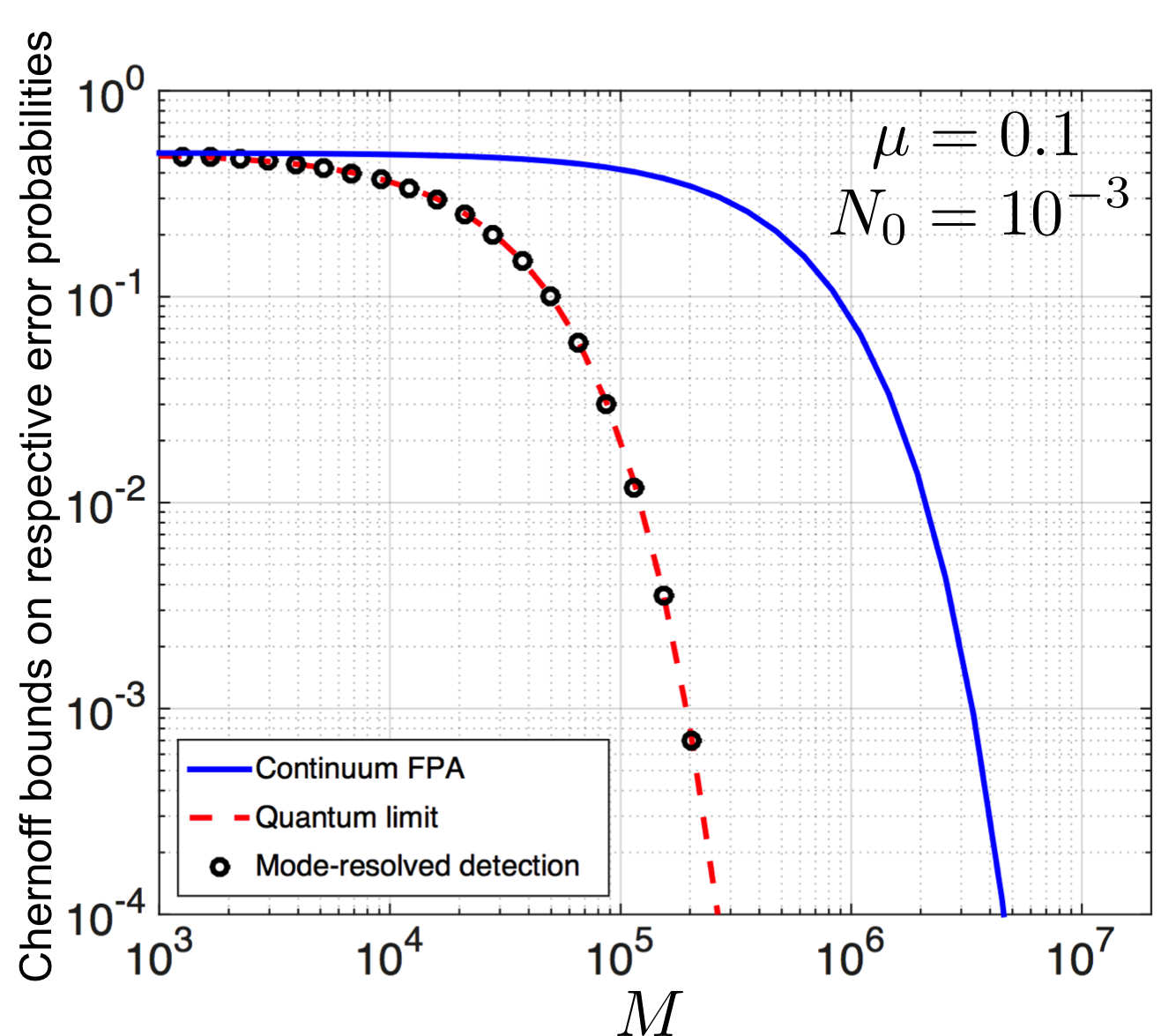}
\caption{For $\mu = 0.1$ and $N_0 = 10^{-3}$ photons per mode, we compare the (classical) Chernoff bound on the error probability attained by a shot-noise-limited continuum-detection focal plane array (solid blue plot) developed in Section~\ref{sec:DD}, the (quantum) Chernoff bound on the minimum error probability attainable by the optimal detector (red dashed plot) developed in Section~\ref{sec:quantum}, and the (classical) Chernoff bound on the error probability attained by the spatial-mode-resolved $3$-pixel detector (black circles) developed in Section~\ref{sec:structured}.} 
\label{fig:QCB_CCB_compare}
\end{figure}

In Ref.~\cite{2008PirandolaLloyd}, it was shown how to calculate the Chernoff bound for discriminating $\rho_0^{\otimes M}$ and $\rho_1^{\otimes M}$, where $\rho_0$ and $\rho_1$ are $n$-mode Gaussian states, using the symplectic eigenvalues and eigenvectors of the respective covariance matrices ${\boldsymbol V}^{(0)}$ and ${\boldsymbol V}^{(1)}$. The quantum Chernoff exponent is $-\log Q$ with $Q \equiv \min\limits_{0 \le s \le 1} Q(s)$, where $Q(s) = {\rm Tr}\left(\rho_0^s\rho_1^{1-s}\right)$. The expression for $Q(s)$ for the case of $n$-mode zero-mean states $\rho_0$ and $\rho_1$ are given by~\cite{2008PirandolaLloyd}:
\[
Q(s) = \frac{2^n\prod_{k=1}^nG_s(\alpha_k)G_{1-s}(\beta_k)}{\sqrt{\text{det}[\textbf{V}^{(0)}(s)+\textbf{V}^{(1)}(1-s)]}}\,,
\]
where,
\begin{align}
&\textbf{V}^{(0)}(s)=\textbf{S}_0 \left[\bigoplus_{k=1}^n \Lambda_s(\alpha_k)\textbf{I}_k \right]\textbf{S}_0^T,\nonumber \\
&\textbf{V}^{(1)}(1-s)=\textbf{S}_1 \left[\bigoplus_{k=1}^n \Lambda_{1-s}(\beta_k)\textbf{I}_k \right]\textbf{S}_1^T, \nonumber \\
&G_p(x)=\frac{2^p}{(x+1)^p-(x-1)^p}, {\text{and}}\nonumber \\
&\Lambda_p(x)=\frac{(x+1)^p+(x-1)^p}{(x+1)^p-(x-1)^p}\,.
\end{align}
In the above, the symplectic matrices $\textbf{S}_0$ and $\textbf{S}_1$ diagonalize the covariance matrices $\textbf{V}^{(0)}$ and $\textbf{V}^{(1)}$. For our problem, these quantities can be evaluated to be:
\be
\textbf{S}_0=\textbf{I} \quad \text{and }\quad  \textbf{S}_1=\mx{a&0&-b\\0&1&1\\b&0&a}\,,
\ee
where,
\begin{align}
&\alpha_1=\frac{2N_0+1}{4},\quad \alpha_2=\frac{1}{4},\quad\alpha_3=\frac{1}{4},\nonumber \\
&\beta_1=\frac{2N_0(1+\sinctmu)+1}{4},\quad\beta_2=\frac{2N_0(1-\sinctmu)+1}{4}, \nonumber \\
&\beta_3=\frac{1}{4}, {\text{and}}\nonumber \\
&a=\frac{\sqrt{2}~\sincmu}{\sqrt{1+\sinctmu}}, \quad b=\frac{\sqrt{1+\sinctmu-2\sinc^2 \mu}}{\sqrt{1+\sinctmu}}.\label{eq:ab}
\end{align}
Using these quantities we evaluate $Q_\mu(s)$ as a function of $s$ and numerically determine its minimum to compute the quantum Chernoff exponent $\xi_Q = -\log Q_\mu$, which we then insert in Eq.~\ref{eq:QCB} to evaluate the QCB. In Fig.~\ref{fig:QCB_CCB_compare}, we plot the quantum Chernoff bound on the minimum error probability (attained by an optimal detector with unspecified structured realization) and the classical Chernoff bound on the error probability attained by an ideal shot-noise-limited continuum image plane detector, evaluated in Section~\ref{sec:DD}. We used the same parameters as used in Fig.~\ref{fig:FPA_pixelated_Pe} ($\mu = 0.1$ and $N_0 = 10^{-3}$ photons per mode) in order to make the comparison easier. Fig.~\ref{fig:QCB_CCB_compare} also shows the (classical) Chernoff bound on the error probability attained by structured receiver, that we propose and analyze next in Section~\ref{sec:structured}, which is seen to attain quantum-limited performance. In the next Section, we will show that this structured receiver we propose in Section~\ref{sec:structured}, which employs pre-detection spatial mode sorting followed by detection using three temporal-mode-resolved single-photon-sensitive detectors, exactly achieves the quantum limited performance for all values of angular separation $\mu$, and is always significantly superior to the performance attainable by an ideal continuum image-plane detector.

\section{Structured receiver using mode-resolved photon detection}\label{sec:structured_main}

\subsection{Structured receiver}\label{sec:structured}
Let us recall that the span $\cal S$ of the three spatial modes in Eqs.~\eqref{eq:aperturemode1},~\eqref{eq:aperturemode2} and~\eqref{eq:aperturemode3} contain all the relevant information in this imaging problem. Let us consider a receiver that extracts these specific orthogonal spatial modes $\phi_1(x^\prime), \phi_2(x^\prime)$ and $\phi_3(x^\prime)$ (over all $M$ statistically-independent temporal modes) using a tailor-made mode sorter, followed by temporal-mode-resolved photon detection on each of the $M$ orthogonal temporal modes of the three separated spatial modes. For the ensuing analysis, we will assume that each of the $3M$ orthogonal spatio-temporal modes are detected by unity-efficiency shot-noise-limited photon-number resolving detectors. However, photon number resolution in these detectors is unnecessary, and the difference in performance from using single-photon detectors would not be perceivable, since the mean photon number per mode $N_0 \ll 1$. There have been recent proposals on temporal-mode-resolved single photon detection using sum-frequency generation in nonlinear waveguides~\cite{Kan16}. The receiver obtains a vector of detected-photon-number outcomes over all the $3M$ modes, and using that makes a guess on which hypothesis $H_0$ or $H_1$ is true. The probability mass function of $(k_1, k_2, k_3$, the click record for the three spatial modes in one of the $M$ temporal modes is given by:
\be
P^{(0)}(k_1, k_2, k_3) = \frac{N_0^{k_1}}{(1+N_0)^{1+k_1}}\,\delta_{k_2,0}\delta_{k_3,0},
\ee
under hypothesis $H_0$. Under hypothesis $H_1$, the joint probability mass function of $(k_1, k_2, k_3$ is given by
\begin{eqnarray}
P^{(1)}(k_1, k_2, k_3) &=& \frac{N_1^{k_1+k_3}}{(1+N_1)^{1+k_1+k_3}}\binom{k_1+k_3}{k_1}\,\eta^{k_1}(1-\eta)^{k_3} \nonumber \\
&& \times \frac{N_2^{k_2}}{(1+N_2)^{1+k_2}},
\end{eqnarray}
where $N_1 = N_0\left(1 + {\rm sinc}(2\mu)\right)/2$, $N_2 = N_0 - N_1 = N_0\left(1 - {\rm sinc}(2\mu)\right)/2$, and $\eta = 2{\rm sinc}^2(\mu)/(1+{\rm sinc}(2\mu))$. Note that $\eta = a^2 = 1 - b^2$ where $a$ and $b$ are defined in Eq.~\eqref{eq:ab}. Since $N_0 \ll 1$, almost all the photon count records $k_i^{(m)}$, $m = 1, \ldots, M$, and $i = 1, 2, 3$ are either $0$ or $1$. Hence, no perceivable receiver performance loss will be seen if the detectors are on-off single-photon-sensitive detectors that tell whether or not a given spatio-temporal mode has a photon or not.

The probability of error achieved by this receiver is upper bounded by the classical Chernoff bound (see Appendix~\ref{app:CB}) as:
\be
P_{R} \le \frac{1}{2}e^{-M \xi_R},
\ee
with $\xi_R = -\log(S_\mu)$, where 
\be
S_\mu = \min\limits_{0 \le s \le 1} \sum_{k_1, k_2, k_3} \left[P^{(0)}(k_1, k_2, k_3)\right]^s\left[P^{(1)}(k_1, k_2, k_3)\right]^{1-s}. \nonumber
\ee
We can explicitly evaluate $S_\mu$ as a function of $\mu$ and $N_0$:
\begin{eqnarray}
S_\mu(\mu,N_0) &=& \frac{2}{\left(N_0(1-{\rm sinc}(2\mu)) + 2\right)} \nonumber \\
&&\times \frac{2}{\left(N_0(1+{\rm sinc}(2\mu) - 2{\rm sinc}^2\mu)+ 2\right)}.
\end{eqnarray}
Using this expression, we evaluate and plot the (classical) Chernoff bound of the receiver as shown in Fig.~\ref{fig:QCB_CCB_compare}. We numerically find that the Chernoff exponent of our receiver $\xi_R$ equals the quantum Chernoff exponent $\xi_Q$ (see Eq.~\eqref{eq:QCB}) exactly for all values of $N_0$ and $\mu$. 

\subsection{Performance comparison}

\begin{figure}
\centering
\includegraphics[width=\columnwidth]{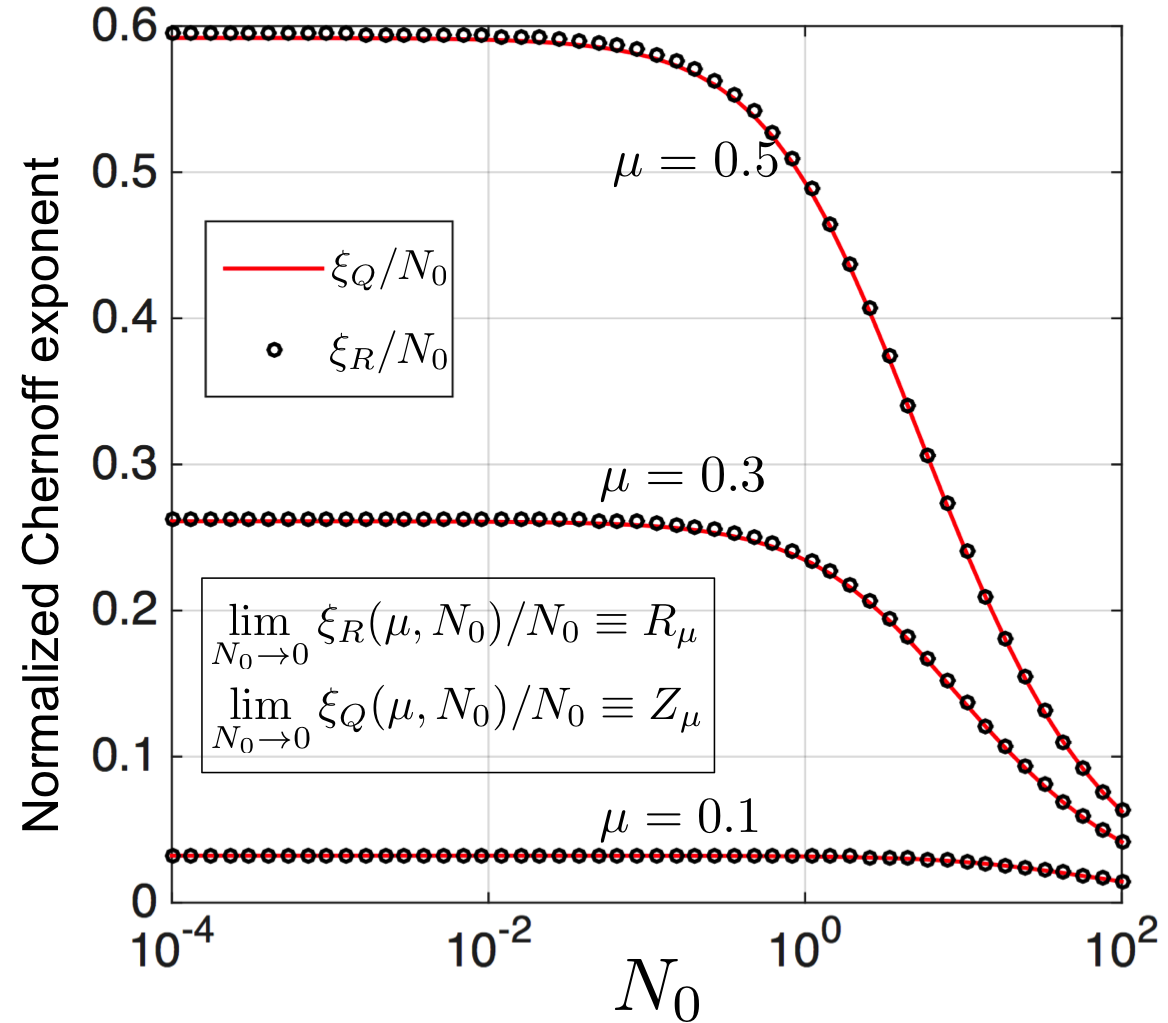}
\caption{We plot the (quantum) Chernoff exponent of the optimal receiver $\xi_Q(\mu,N_0)$ and the (classical) Chernoff exponent of the mode-resolved-detection receiver $\xi_R(\mu,N_0)$, normalized by $N_0$. We see that when $N_0 \ll 1$, the ratios $\xi_R(\mu,N_0)/N_0$ and $\xi_Q(\mu,N_0)/N_0$ are equal and they converge to an $N_0$-independent values that only depends on $\mu$. We call these limiting values of the ratios $R_\mu$ and $Z_\mu$ respectively.}
\label{fig:normalizedCE}
\end{figure}
\begin{figure}
\centering
\includegraphics[width=0.9\columnwidth]{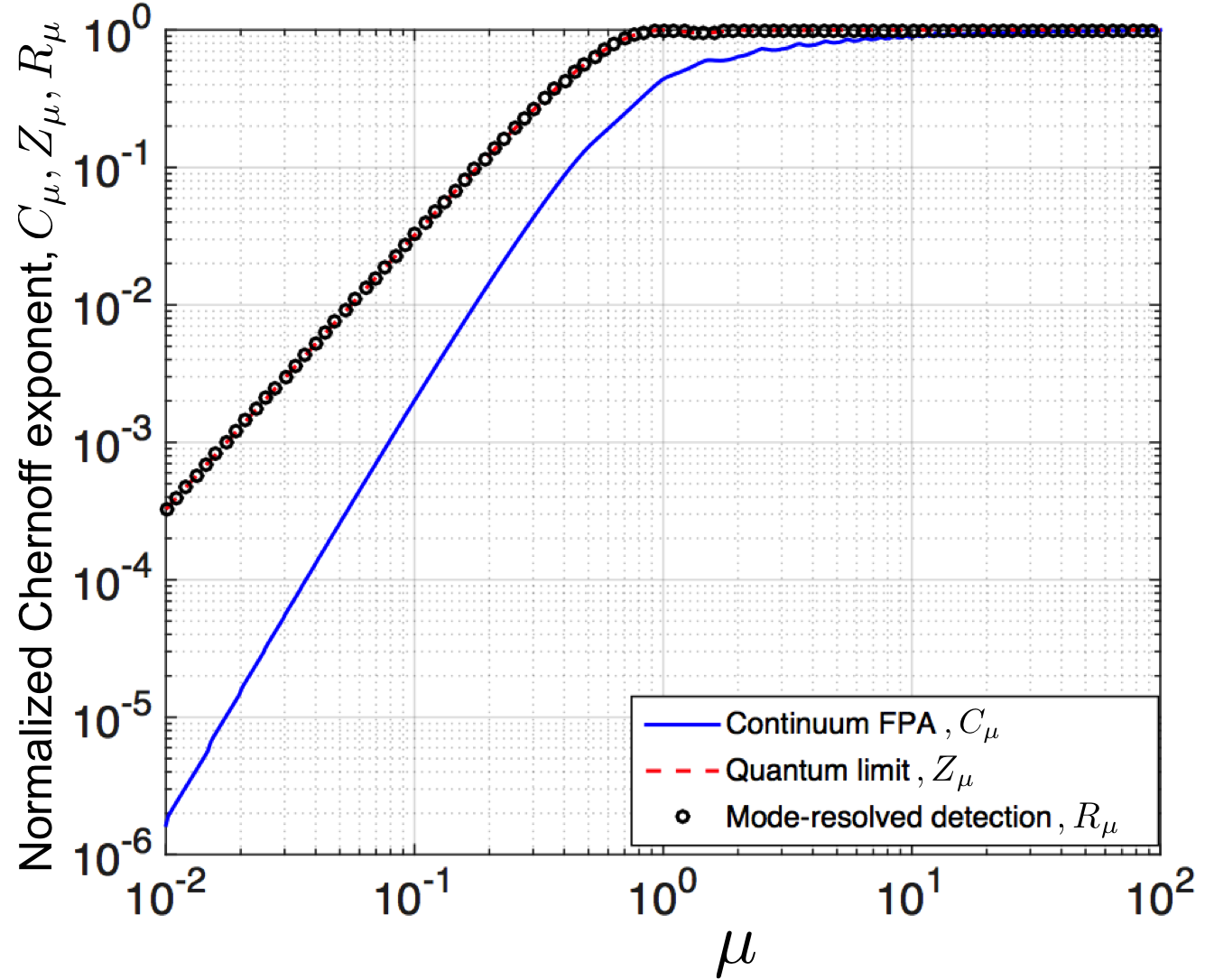}
\caption{Plot of the normalized error exponents $C_\mu$ (ideal continuum FPA) evaluated in Section~\ref{sec:DD}, $Z_\mu$ (quantum-optimal receiver) evaluated in Section~\ref{sec:quantum} and $R_\mu$ (mode-resolved three-pixel structured receiver) evaluated in Section~\ref{sec:structured} plotted as a function of $\mu$. The probability or error of the respective receiver is upper bounded by $\frac{1}{2}\exp(-NV_\mu)$, where $V_\mu$ is the respective receiver's exponent ($C_\mu, Z_\mu$ or $R_\mu$) and $N = N_0M$ ($M \approx WT$) is the total mean photon number collected during the receiver's integration time $T$.}
\label{fig:exp_compare}
\end{figure}

In order to compare the Chernoff bounds for the three cases we have studied for far, we will first argue that when $N_0 \ll 1$, for each of the three cases, the respective (classical or quantum) Chernoff bound can be expressed as $\frac{1}{2}e^{-NX_\mu}$, where $N = MN_0$ is the total photon number collected over the integration time ($M$ temporal modes), and the normalized Chernoff exponent $X_\mu$ is only a function of $\mu$, and not a function of $N_0$. As discussed in Section~\ref{sec:DD}, the continuum detector's classical Chernoff bound already (exactly) has this feature for any $N_0$, i.e., $P_{\rm DD} \le \frac{1}{2}e^{-NC_\mu}$. The minimum error probability of the optimal detector satisfies the (quantum) Chernoff bound $P_{\rm min} \le \frac{1}{2}e^{-M\xi_Q(\mu,N_0)}$, and the error probability achieved by the mode-resolved detector discussed above satisfies the (classical) Chernoff bound $P_{R} \le \frac{1}{2}e^{-M\xi_R(\mu,N_0)}$. In Fig.~\ref{fig:normalizedCE}, we plot the ratios $\xi_Q(\mu,N_0)/N_0$ and $\xi_R(\mu,N_0)/N_0$ as a function of $N_0$, and find that each converges to a respective $N_0$-independent constant that only depends on $\mu$. We name these two constants $\lim_{N_0 \to 0}\xi_Q(\mu,N_0)/N_0 \equiv Z_\mu$, and $\lim_{N_0 \to 0}\xi_R(\mu,N_0)/N_0 \equiv R_\mu$ respectively. In terms of these newly-defined normalized Chernoff exponents, we now have that, for $N_0 \ll 1$, $P_{\rm min} \le \frac{1}{2}e^{-NZ_\mu}$ and $P_{R} \le \frac{1}{2}e^{-NR_\mu}$, where $N$ is the total photon number collected over the integration time. Now that we have all three Chernoff exponents normalized and completely expressed only in terms of the half-angular separation $\mu$, we can compare them with each other as a function of $\mu$. We do this in Fig.~\ref{fig:exp_compare}.

We will now argue that when $\mu \to \infty$ (the sources are far separated), the error probability achieved by each of the three detectors we have discussed in this paper goes to $\frac{1}{2}e^{-N}$. First, the argument for the image-plane continuum direct detection goes as follows. In the $\mu \to \infty$ limit, the image-plane intensity pattern ($I_0(x)$ or $I_1(x)$) for each of the two hypotheses can be attributed to a set of non-overlapping detector pixels in the FPA. So, the probability of making an error given either hypothesis is true is the probability that the $N$-photon waveform $I_0(x)$ or $I_1(x)$ does not generate a single click in its own block of detector pixels, which is $e^{-N}$, times the probability of getting a random guess correct should a click were to not occur, which happens with probability $1/2$ assuming equally-likely hypotheses. This gives us $P_{\rm DD} = \frac{1}{2}e^{-N}$. Let us now consider the optimal (quantum-limited) performance. The upper panel of Fig.~\ref{fig:intuition} pictorially depicts the states of one temporal mode of the three orthonormal spatial modes of the aperture field $(\phi_1(x^\prime), \phi_2(x^\prime), \phi_3(x^\prime))$. The quantum states of each temporal mode are identical. Under $H_0$, $\phi_1$ is in a zero-mean thermal state with mean photon number $N_0$ whereas $\phi_2$ and $\phi_3$ are in their vacuum states. Under $H_1$, $\phi_2$ is in a zero-mean thermal state---independent of the $(\phi_1, \phi_3)$ joint state---with mean photon number $N_2 = N_0\left(1-{\rm sinc}(2\mu)\right)/2$, and $(\phi_1, \phi_3)$ are in a classically-correlated zero-mean two-mode Gaussian state as described in Section~\ref{sec:quantum} where the mean photon number of $\phi_1$ is $\eta N_1$ and that of $\phi_3$ is $(1-\eta) N_1$, with $N_1 = N_0\left(1+{\rm sinc}(2\mu)\right)/2$ and $\eta = 2{\rm sinc}^2(\mu)/\left[1+{\rm sinc}(2\mu)\right]$. In the $\mu \to \infty$ limit, we have $\eta = 0$, and $N_1 = N_2 = N_0/2$. So, under $H_0$, the joint state of the $(\phi_1, \phi_2, \phi_3)$ modes is $\rho_0 = \sigma_{\rm T}(N_0) \otimes |0\rangle \langle 0| \otimes |0\rangle \langle 0|$, whereas under $H_1$, the joint state is $\rho_1 = |0\rangle \langle 0| \otimes \sigma_{\rm T}(N_0/2) \otimes \sigma_{\rm T}(N_0/2)$, where $\sigma_{\rm T}(N_0)$ is a zero-mean thermal state with mean photon number $N_0$. Since both $\rho_0$ and $\rho_1$ are diagonal in the joint photon-number basis, the optimal measurement that achieves the minimum probability of error is photon number basis measurement on all three modes. Given $\sigma_{\rm T}(N_0) = \sum_{n=0}^\infty \left(N_0^n/(1+N_0)^{n+1}\right)|n\rangle \langle n|$, the probability of error is given by $\frac{1}{2}\left(\frac{1}{1+N_0}\right)^{M}$. Substituting $M = N/N_0$, and taking the $N_0 \to 0$ limit, we get $P_{\rm min} = \frac{1}{2}e^{-N}$. Finally, it is simple to see that the mode-resolved detector described in Section~\ref{sec:structured} reduces to photon number measurement on the three modes $(\phi_1, \phi_2, \phi_3)$ in the $\mu \to \infty$ limit, and hence in that limit, $P_{R} = \frac{1}{2}e^{-N}$.

Therefore, $C_\mu$, $R_\mu$ and $Z_\mu$ should each $\to 1$, as $\mu \to \infty$. This is verified in the plots in Fig.~\ref{fig:exp_compare}. It is clear that at all values of $\mu$, our mode-resolved-detection receiver {\em exactly} achieves the QCB, i.e., $Z_\mu = R_\mu$, $\forall \mu$. The convergence of the normalized Chernoff exponent to the large-separation error performance limit of $\frac{1}{2}e^{-N}$ for the continuum FPA (i.e., $C_\mu \to 1$) is seen to be much slower compared to those of the optimal measurement and the mode-resolved detector. This is because the traditional image-plane (pixel-basis) measurement is highly susceptible to errors from the long tails of the sinc-function PSFs of the hard aperture, even when the angular separation is several times the Rayleigh separation $\lambda/D$.


\section{On attaining the quantum limit for general passive imaging problems}\label{sec:general}

\begin{figure}
\centering
\includegraphics[width=\columnwidth]{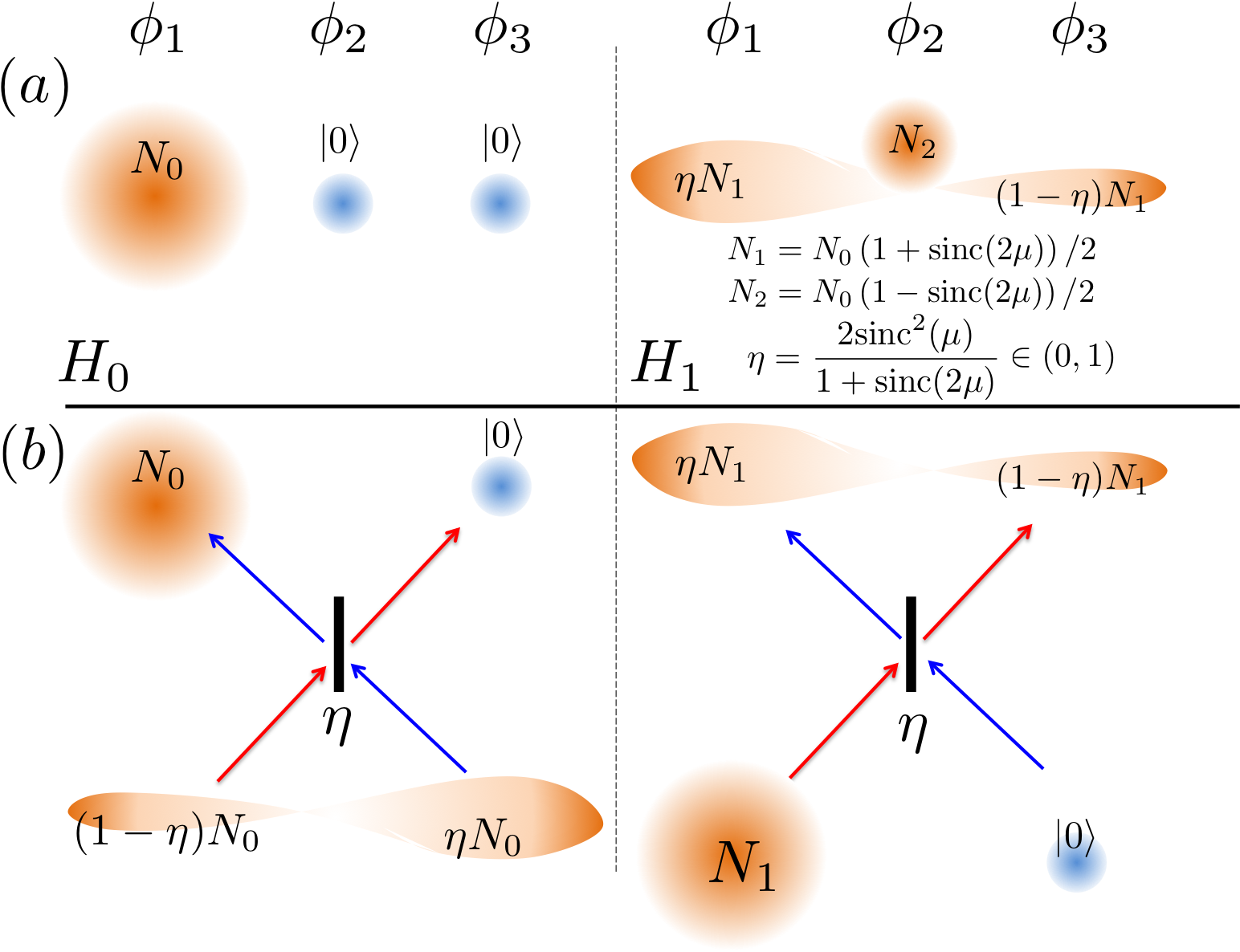}
\caption{(a) A schematic depiction of the quantum states of one temporal mode of the three orthonormal spatial modes of the aperture field $(\phi_1(x^\prime), \phi_2(x^\prime), \phi_3(x^\prime))$; (b) We consider a simplified two-mode multi-copy state discrimination problem where in hypothesis $H_0$ the two modes ($\phi_1$ and $\phi_3$ in this case as shown) are in a product of a thermal state and vacuum, and in hypothesis $H_1$, the two modes are in a classically-correlated two-mode Gaussian state. We show that (see Section~\ref{sec:general} for details) a mixing of the two modes on a beamsplitter of appropriately-chosen transmissivity followed by photon number measurement always achieves quantum-limited minimum error probability performance.}
\label{fig:intuition}
\end{figure}
In Section~\ref{sec:structured_main}, we considered a detection mechanism where we extract a specific set of mutually-orthonormal modes and do quantum-noise-limited intensity measurement on those modes. For the one-vs.-two source hypothesis-test problem we have considered in this paper, we showed that this receiver achieves optimal performance, in the limit of a large number of temporal modes collected over the integration-time source-bandwidth product. The fact that pre-detection optical-domain pre-processing can enhance an imager's performance, and that the pixel-basis measurement (intensity detection in the conventional image plane by a focal plane array) may not be always optimal, have been long known and explored in the computational imaging community, and is usually termed {\em super-resolution}. Recent work by Mankei Tsang and collaborators~\cite{Tsa09,Tsa15,Tsa15b,Nai16,Tsa16_general} have shown that in certain simple imaging problems involving incoherent light, a pre-detection mode sorting can attain quantum limited performance. In particular, Tsang {\em et al.}'s work has established---for the problem of estimating the angular separation between two point sources---that Rayleigh's criterion for imaging resolution is an artifact of the conventional philosophy of focusing the image on a focal plane array and measurement of the field's intensity profile, and that if one allows for a pre-detection mode sorting, the imager's performance (in estimating the angular separation between two point sources) is only a function of the total collected photon energy during the integration time, and is entirely independent of the actual angular separation, no matter how small it is~\cite{Tsa15b}. The above result was proven in the sense of the classical Cramer Rao bound of the mode-resolved detector achieving the quantum Cramer Rao bound. Although mean squared error of an estimate does not always saturate the Cramer Rao bound, the above result is a good evidence that quantum-limited performance for this problem may be exactly attainable by mode-resolved photo detection. 

We conjecture that in a multi-copy estimation or hypothesis test problem in phase-insensitive incoherent-light optical imaging--- multi-copy in the sense of many statistically independent temporal modes with identical mutual coherence functions across all the spatial modes of interest---that a mode-resolved photo detection in at appropriate mode basis always attains the optimal (quantum-limited) performance. 

In an initial attempt to substantiating the aforesaid conjecture, we consider a two-mode multi-copy state discrimination problem inspired by the three-mode multi-copy state discrimination problem considered in this paper. The problem we consider is depicted in Fig.~\ref{fig:intuition}(b). Under Hypothesis $H_0$, two modes are in a product of a thermal state and vacuum, and in hypothesis $H_1$, they are in a classically-correlated two-mode Gaussian state. Under $H_0$, we assume the total photon number across the two modes equals $N_0$ and under $H_1$, we assume the total photon number across the two modes equals $N_1 < N_0$. It is well known that there always exists a mixing ratio $\eta \in (0, 1)$ such that if the two modes are mixed on a beamsplitter of transmissivity $\eta$, that the output of the beamsplitter will be an uncorrelated state, a product of a thermal state and vacuum.

We show that mixing of the two modes on a beamsplitter of an appropriately-chosen transmissivity $\eta$, such that under one of the two hypotheses we have a product of a thermal state and vacuum, followed by photon number measurement in that mode basis, always achieves quantum-limited minimum error probability performance. We show this by showing that the classical Chernoff bound of the output of a particular mode-resolved photon detection always exactly attains the quantum Chernoff bound (for which the measurement is unspecified). 
The derivation of the quantum Chernoff bound for this problem is in Appendix~\ref{app:two mode}.

Connecting the above back to the three-mode problem we studied in this paper, we refer the reader to Fig.~\ref{fig:intuition} again. In our structured receiver design, we did photon-number-basis measurement on the $(\phi_1, \phi_2, \phi_3)$ spatial-mode basis, whose quantum state description is showed in the top panel of Fig.~\ref{fig:intuition}(a). Let us imagine applying a mode mixing (a beamsplitter) of transmissivity $\eta = 2{\rm sinc}^2(\mu)/\left[1+{\rm sinc}(2\mu)\right]$ on the spatial modes $\phi_1$ and $\phi_3$. We would think of the action of that mode mixer alternatively as the extraction of a different orthonormal spatial mode basis of the span $\cal S$. Let us name this new mode basis, $\phi_1^\prime, \phi_2, \phi_3^\prime$. The state of $\phi_2$ remains same as above (vacuum for $H_0$ and $\sigma_T(N_2)$ for $H_1$, uncorrelated with the remaining two modes) but the joint state of $(\phi_1^\prime, \phi_3^\prime)$ is now different, and depicted at the bottom panel of Fig.~\ref{fig:intuition}(b). In the original $(\phi_1, \phi_2, \phi_3)$ spatial-mode basis, the joint state of modes $\phi_1$ and $\phi_3$ is a product state ($\sigma_T(N_0) \otimes |0\rangle \langle 0|$) under $H_0$ and a correlated state under $H_1$. However in the new $(\phi_1^\prime, \phi_2, \phi_3^\prime)$ basis, the joint state of modes $\phi_1^\prime$ and $\phi_3^\prime$ is a correlated state under $H_0$ and a product state ($\sigma_T(N_1) \otimes |0\rangle \langle 0|$) under $H_1$. The performance of a measurement that Helstrom evaluated~\cite{Hel73} was done in the $(\phi_1^\prime, \phi_2, \phi_3^\prime)$ mode basis, where mode $\phi_2$ was detected (in the photon number basis) first, followed by the quantum-optimal measurement (with unspecified physical realization) on the joint-state of the modes $(\phi_1^\prime, \phi_3^\prime)$. Since Helstrom only considered an optimal measurement and not a structured receiver realization on the joint state of the $(\phi_1^\prime, \phi_3^\prime)$ modes, there would have had been no difference in his calculations if he instead chose to evaluate his measurement's performance on the joint state of the $(\phi_1, \phi_3)$ modes instead. The reason for this is that the action of the beamsplitter on the quantum state of the two modes is a unitary, which does not change the performance of the optimal quantum-limited measurement on those. However, our result in this paper shows that quantum-limited direct detection in the $(\phi_1, \phi_2, \phi_3)$ mode basis {\em exactly} attains quantum-optimal performance. This means that after the direct detection of the $\phi_2$ mode, had Helstrom chosen to evaluate the performance of a direct detection measurement in his $(\phi_1^\prime, \phi_2, \phi_3^\prime)$ mode basis, then it would have had suboptimal performance. 

Even if one proves the conjecture---that mode-resolved photo detection in at appropriate mode basis always attains the quantum optimal performance for all incoherent optical imaging problems involving multimode phase-insensitive thermal light (a problem also alluded to and analyzed recently by Tsang in~\cite{Tsa16_general})---understanding the optimal choice of the mode basis in the context of a given problem, and more importantly, understanding structured and systematic means of physically realizing arbitrarily-programmable spatio-temporal mode transformations in a low-loss fashion will be instrumental to reaping these benefits of non-standard detection methods in optical imaging.

\section{Conclusion}\label{sec:conclusion}

We considered the problem of discriminating one from two closely-spaced incoherently-radiating quasi-monochromatic point sources. We assumed that the total photons collected over the integration time is the same under either hypothesis, so that the total measured intensity bears no signature of the hypothesis. For the one-source hypothesis, the point source was taken to be on-axis along the line of sight, and for the two-source hypothesis, we gave ourselves the prior knowledge of the angular separation of the sources, and were assumed to be located symmetrically off-axis. This problem was studied by Helstrom in 1973, who evaluated the probability of error achievable using a sub-optimal optical measurement, with an unspecified structured realization. We evaluated the quantum Chernoff bound, an upper bound on the minimum probability of error achievable by any physically-realizable receiver, which is asymptotically tight in the regime that the integration time is high. We evaluated the classical Chernoff bound on the error probability achievable using an ideal image plane array---a signal shot-noise limited continuum photon-detection receiver with infinitely many infinitesimally-tiny pixels---and quantified its performance gap with the quantum limit. Finally, we showed that an explicit structured receiver that separates three orthogonal spatial modes of the aperture field followed by detecting them using three temporally mode-resolved single-photon-sensitive detectors, exactly achieves the quantum-limited performance. In other words, the classical Chernoff bound of the three-pixel camera with a pre-detection mode sorter exactly matches with the quantum Chernoff bound for this problem. We discussed why we believe the above observation---that an appropriate pre-detection mode-sorting followed by shot-noise-limited photon detection is the optimal detection technique---is not unique for this particular problem we studied, and is generally true for all imaging problems where a scene parameter is encoded in incoherent light.

\acknowledgements

SG would like to acknowledge helpful discussions with Mankei Tsang and Christopher Fuchs. The authors thank Mankei Tsang for pointing to related work by their group while this paper was being written, which also addresses the one-vs.-two incoherent point-source discrimination problem~\cite{Tsa16a}. HK and SG were supported in part by the DARPA Information in a Photon (InPho) program under prime contract number HR0011-10-C-0159, and the DARPA Revolutionary Enhancement of Visibility by Exploiting Active Light-fields (REVEAL) program under a subcontract to University of Arizona Tucson with prime contract number HR0011-16-C-0026. JHS is supported in part by the DARPA REVEAL program.

\appendix

\section{Sampling from discrete point processes}\label{app:CB}

Consider a test for equally-likely binary hypotheses in which we observe a point process on $-\infty < x < \infty$ that is Poisson with rate function $\lambda_j(x)$, under hypothesis $H_j$, where $\int_{-\infty}^\infty\!{\rm d}x\,\lambda_j(x) < \infty$, for $j=0,1$.  Let $0 \le N < \infty$ be the total number of observed occurrences of this point process, and let $\{\,x_n : 1 \le n \le N\,\}$ be its observed occurrences.  When a minimum error-probability decision as to the true hypothesis is made, based on observation of the point process,  the resulting error probability satisfies the Chernoff bound \cite{Snyder},
\begin{align}
&\Pr(e)_{\rm min} \le \min_{0\le s \le 1}\exp(\int_{-\infty}^\infty\!{\rm d}x\,[-s\lambda_0 (x) - (1-s)\lambda_1(x) \nonumber \\ 
&+\lambda_0^s(x)\lambda_1^{(1-s)}(x)])/2.
\end{align} 

Now suppose we discretize space into pixels of length $\Delta > 0$, i.e., instead of observing $\{\,x_n : 1 \le n \le N\,\}$ and using that data in our hypothesis test, we observe $\{\,y_k : -\infty < k < \infty\,\}$, where
\begin{equation}
y_k = \int_{(k-1/2)\Delta}^{(k+1/2)\Delta}\!{\rm d}x\,\sum_{n=1}^N \delta(x-x_n),
\end{equation}
for $\delta(\cdot)$ being the Dirac delta function.  Under hypothesis $H_j$, the $\{y_k\}$ are thus independent, Poisson random variables with mean values
\begin{equation}
\mu_j[k] = E[y_k\mid H_j] = \int_{(k-1/2)\Delta}^{(k+1/2)\Delta}\!{\rm d}x\,\lambda_j(x).
\end{equation}
The Chernoff bound on the minimum error-probability test using the pixellated data is 
\begin{align}
&2\Pr(e)_{\rm min} \le\nonumber \\
&\min_{0\le s\le 1}\prod_{k=-\infty}^\infty \left[\sum_{n=0}^{\infty}\left(\frac{\mu_0^n[k]e^{-\mu_0[k]}}{n!}\right)^s
\left(\frac{\mu_1^n[k]e^{-\mu_1[k]}}{n!}\right)^{(1-s)}\right] \nonumber \\
&= \min_{0\le s\le 1}\prod_{k=-\infty}^\infty \exp\Big(-s\mu_0[k] -(1-s)\mu_1[k] + \nonumber \\
&\qquad\qquad \qquad \mu_0^s[k]\mu_1^{(1-s)}[k]\Big) \nonumber\\
&= \min_{0\le s \le 1}\exp\Big[-\sum_{k=-\infty}^\infty\Big(s\mu_0[k] +(1-s)\mu_1[k] - \nonumber \\
&\qquad\qquad\qquad\mu_0^s[k]\mu_1^{(1-s)}[k]\Big)\Big]. 
\end{align}

\section{Two mode problem}\label{app:two mode}

Formally, the problem we consider is to discriminate between $M$ i.i.d. copies of the state $\rho_{1}\otimes \ket{0}\bra{0}$ and the state $\rho_{2}\otimes \ket{0}\bra{0}$ after it passes through a beamsplitter of transmissivity $\eta$, where $\rho_{i}$ is a thermal state of mean photon number $N_i$. In other words, the covariance matrices are of the form
\be
\textbf{V}^{(1)}_A=\mx{\frac{2N_1+1}{4}&0\\0&\frac{1}{4}} \quad\text{and}\quad \textbf{V}^{(1)}_B=\textbf{S}_B\mx{\frac{2N_2+1}{4}&0\\0&\frac{1}{4}}\textbf{S}_B^T\,,
\ee
where
\be
\textbf{S}_B=\mx{\sqrt{\eta}&\sqrt{1-\eta}\\-\sqrt{1-\eta}&\sqrt{\eta}}\,.
\ee
The expressions needed for the Chernoff bound in this case are
\begin{align}
&\alpha_1=\frac{2N_1+1}{4}\quad \text{and}\quad \alpha_2=\frac{1}{4}\nonumber \\
&\beta_1=\frac{2N_2+1}{4}\quad \text{and}\quad \beta_2=\frac{1}{4}\,.
\end{align}
Using these quantities, the quantum Chernoff bound is evaluated and compared to the classical Chernoff bound for direct detection.

\bibliographystyle{apsrev4-1}
\bibliography{imaging} 

\end{document}